\documentclass[conference]{IEEEtran}
\IEEEoverridecommandlockouts
\usepackage{cite}
\usepackage{amsmath,amssymb,amsfonts}
\usepackage{algorithmic}
\usepackage{graphicx}
\usepackage{tabularx}
\usepackage{textcomp}
\usepackage{lipsum}
\usepackage{xcolor}
\def\BibTeX{{\rm B\kern-.05em{\sc i\kern-.025em b}\kern-.08em
    T\kern-.1667em\lower.7ex\hbox{E}\kern-.125emX}}
\begin{document}

\title{OKSP: A Novel Deep Learning Automatic Event Detection Pipeline for Seismic Monitoring \\ in Costa Rica\\}

\author{%
\IEEEauthorblockN{Leonardo van der Laat}
\IEEEauthorblockA{\textit{Costa Rica High Technology} \\
\textit{Center}\\
San José, Costa Rica \\
leonardo.vanderlaat.munoz@una.\\
ac.cr
}
\and
\IEEEauthorblockN{Ronald J.L. Baldares}
\IEEEauthorblockA{\textit{Costa Rica High Technology} \\
\textit{Center and}\\
\textit{Costa Rica Institute of}\\
\textit{Technology}\\
San José, Costa Rica \\
rloaizab@cenat.ac.cr}
\and
\IEEEauthorblockN{Esteban J. Chaves}
\IEEEauthorblockA{\textit{Volcanological and} \\
\textit{Seismological Observatory of}\\
\textit{Costa Rica}\\
\textit{National University}\\
Heredia, Costa Rica \\
esteban.j.chaves@una.ac.cr}
\and
\IEEEauthorblockN{Esteban Meneses}
\IEEEauthorblockA{\textit{Costa Rica High Technology} \\
\textit{Center and}\\
\textit{Costa Rica Institute of}\\
\textit{Technology}\\
San José, Costa Rica \\
esteban.meneses@acm.org}

}


\maketitle

\begin{abstract}
Small magnitude earthquakes are the most abundant but the most difficult to locate robustly and well due to their low amplitudes and high frequencies usually obscured by heterogeneous noise sources. They highlight crucial information about the stress state and the spatio-temporal behavior of fault systems during the earthquake cycle, therefore, its full characterization is then crucial for improving earthquake hazard assessment. Modern deep learning algorithms along with the increasing computational power and efficiency are exploiting the continuously growing seismological databases, worldwide, allowing scientists to improve the completeness for earthquake catalogs, systematically detecting and locating smaller magnitude earthquakes and reducing the errors introduced mainly by human intervention through traditional approaches in seismological observatories. In this work, we introduce OKSP, a novel deep learning automatic earthquake detection pipeline for seismic monitoring in Costa Rica. Using Kabré supercomputer from the Costa Rica High Technology Center, we applied OKSP to the day before and the first 5 days following the Puerto Armuelles, M6.5, earthquake that occurred on 26 June, 2019, along the Costa Rica-Panama border and found 1100 more earthquakes previously unidentified by the Volcanological and Seismological Observatory of Costa Rica. From these events, a total of 23 earthquakes with magnitudes below 1.0 occurred a day to hours prior to the mainshock, shedding light about the rupture initiation and earthquake interaction leading to the occurrence of this productive seismic sequence.  Our observations show that for the study period, the model was 100\% exhaustive and 82\% precise, resulting in an F1 score of ~0.90. This effort represents the very first attempt for automatically detecting earthquakes in Costa Rica using deep learning methods and demonstrates that, in the near future, earthquake monitoring routines will be carried out entirely by AI algorithms. 

\end{abstract}


\section{Introduction}

Heterogeneous faults often exhibit an intermingling of fast (seismic) and slow (aseismic) slip, that strongly influences seismic behavior and may promote the nucleation of potentially catastrophic events. Understanding how seismic and aseismic processes communicate as well as the spatio-temporal evolution of complete earthquake sequences during the earthquake cycle is then crucial for improving hazard assessment \cite{Lay2012, Chaves2020}.

\begin{figure*}[bhtp]
    \centerline{\includegraphics[width=0.85\paperwidth]{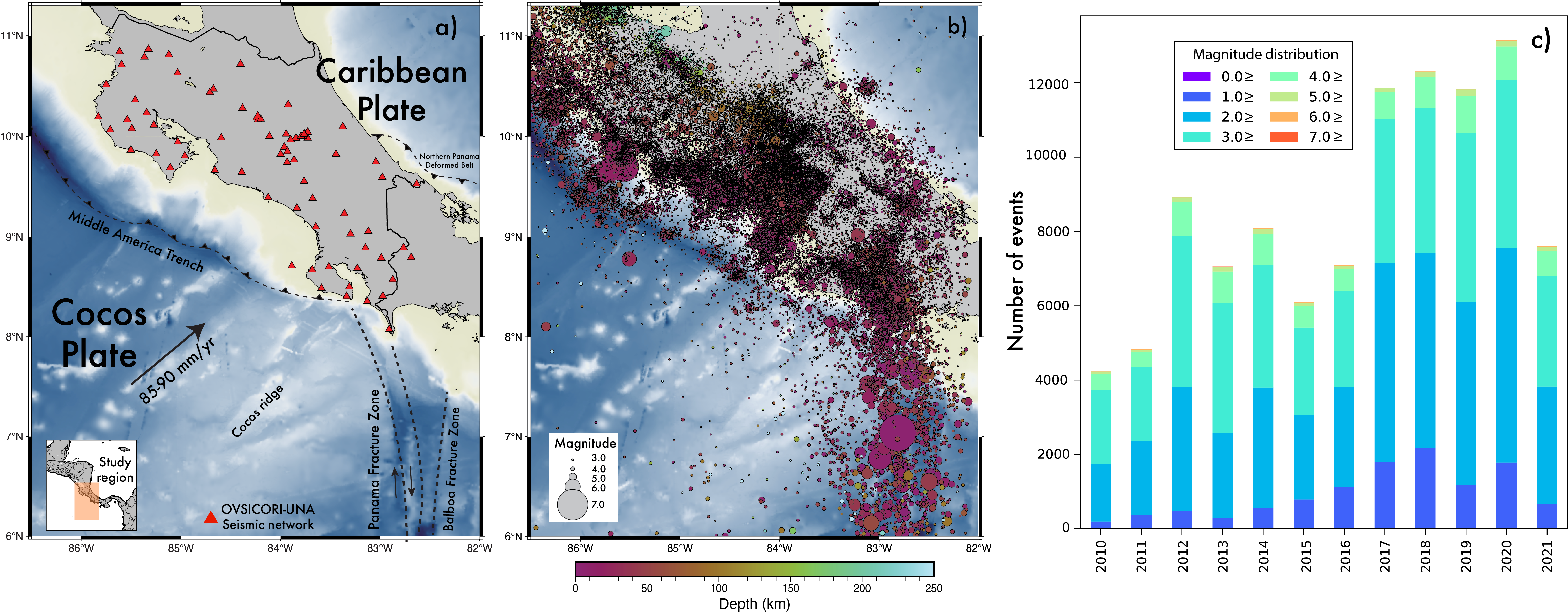}}
    \caption{\textbf{Figure showing the spatial distribution of broadband seismic stations and seismicity of Costa Rica}. Panel a) displays the location of 113 seismological stations (red triangles) distributed along Costa Rica and operated by the Volcanological and Seismological Observatory of Costa Rica at Universidad Nacional (OVSICORI-UNA). The seismic catalog of Costa Rica from 2010, when the digital era initiated in the Observatory, to July 2021, is presented in panel b), where the color and size of the circles represent the earthquake depth and magnitude, respectively. The seismological catalog was created by traditional methods that involve manually and individually identifying, picking, and locating earthquakes. The magnitude distribution and number of events per year is presented in panel c). The color of the bars indicates magnitude range. 
}
    \label{fig:fig1}
\end{figure*}

\begin{figure}[bhtp]
    \centerline{\includegraphics[width=\columnwidth]{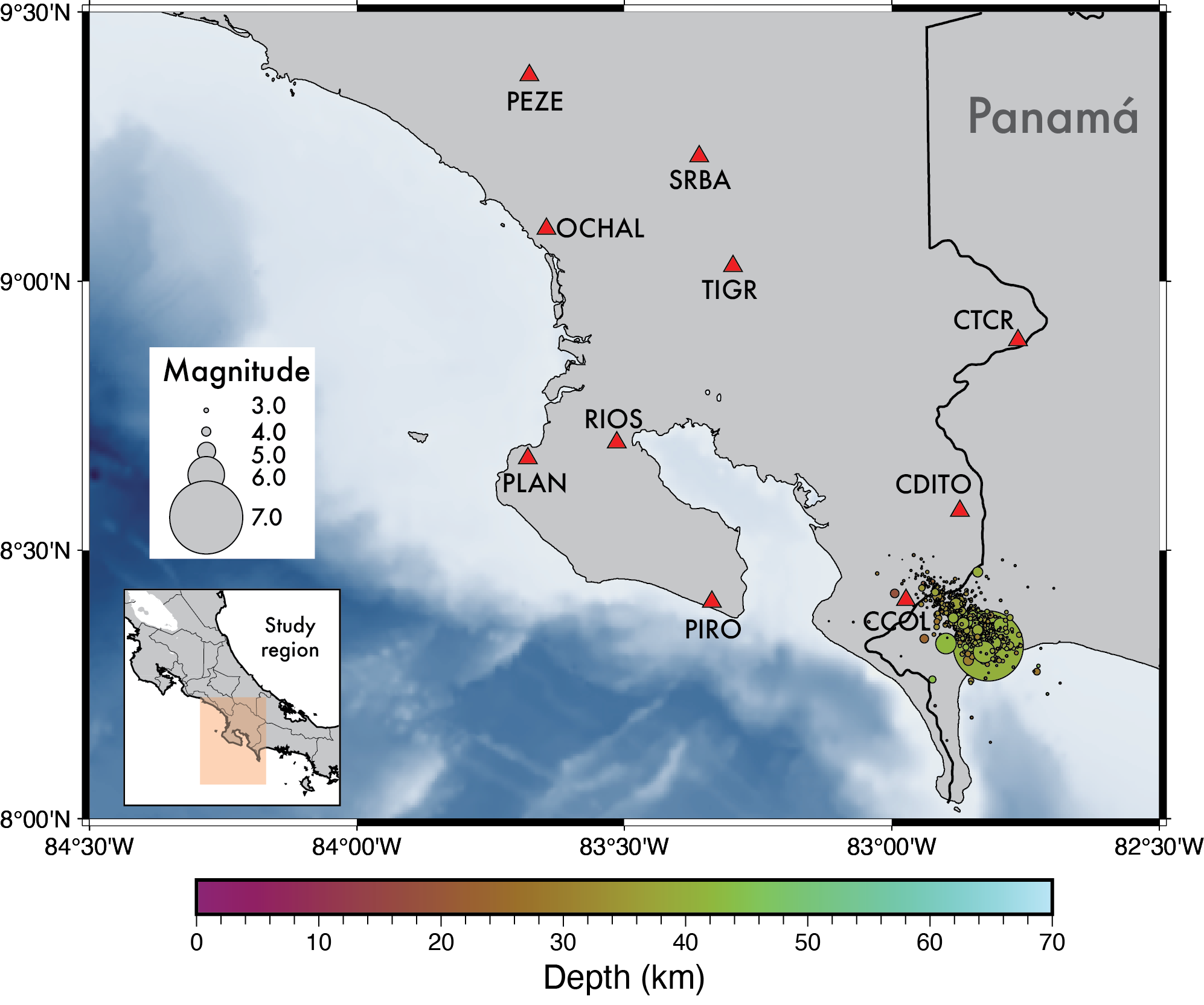}}
    \caption{\textbf{The Puerto Armuelles earthquake sequence and broadband seismic stations used in this study}. Map showing the spatial distribution of the seismicity (circles) located by OVSICORI-UNA from June 25 to June 30, 2019, along the Burica Peninsula, including the Puerto Armuelles mainshock, Mw=6.5, occurred on June 26 at 05:23:47 UTC. The color and size of the circles correspond with the depth and magnitude of the earthquake, respectively. The red triangles represent the seismological stations used in this study. 
}
    \label{fig:fig2}
\end{figure}

The exponentially increasing number and quality of the seismological instrumentation and data worldwide, is continuously improving our understanding of fault physics, allowing seismologists to better characterize the rupture initiation and dynamics of large events and its associated foreshock and aftershock sequences. The expansion of modern seismic networks is also contributing to reduce the magnitude of completeness for modern earthquake catalogs, increasing the number of small magnitude earthquakes \cite{Yoon, ZhuBeroza2018, Ross} that pinpoint the regions of both, weak and strong seismic coupling along the plate interface.

However, the inherent cost of maintaining dense seismic networks is based on the loss, mainly due to human bias, of a significant number of small magnitude earthquakes in the catalog, crucial to fully understand fault behavior during the earthquake cycle \cite{Brodsky}. These events radiate energy with low amplitudes and frequencies that are generally obscured by ambient noise, anthropogenic or/and atmospheric sources, e.g. \cite{Seydoux2020, Lecocq}. Therefore, traditional seismological methods for earthquake detection and location, like STA/LTA \cite{Allen1978} are limiting geophysical observations and may be subject to increasing bias and errors as network instrumentation expands, especially in seismological observatories where multiple analysts are responsible for phase picking (the measurement of arrival times of distinct seismic phases, like P and S phases) and earthquake location.

Advanced machine learning and deep learning algorithms, along with the continuously increasing computer power, have demonstrated to provide a unique opportunity for overcoming those limitations, establishing a state-of-the-art framework that allows to recognize patterns and data characteristics that otherwise would not be possible to observe by usual means~\cite{PulliDysart1990,MusilPlesinger1996,Kong2018,Seydoux2020}. All these algorithms require a substantial amount of computing cycles.

A great opportunity arises when a big seismological data source is coupled with large computing capabilities. Therefore, the Volcanological and Seismological Observatory of Costa Rica at Universidad Nacional (OVSICORI-UNA) teamed up with the Costa Rica National High Technology Center (CeNAT). While the former provides an extensive seismological database (Fig.1), the latter operates Kabré, a general purpose supercomputer. Together, we developed OKSP (OVSICORI Kabré Seismological Pipeline), a novel deep learning pipeline, written in Python, that incorporates the use of EQTransformer~\cite{Mousavi2020} and the seismic network operated by OVSICORI-UNA. OKSP provides automatic event detection, phase picking and association of earthquake seismicity in Costa Rica with the aim of improving catalog completeness to better characterize the local and regional seismotectonics and earthquake potential along fault systems (Fig.\ref{fig:fig1}).

We applied our pipeline to the day before and the first 5 days following the Puerto Armuelles, M6.5, earthquake that occurred on 26 June, 2019, along the subducted portion of the Panama Fracture Zone, at the border between Costa Rica and Panama (Fig. \ref{fig:fig2}). This event was widely felt by the population of both countries and was followed by a productive aftershock sequence (more than 5000 earthquakes during the span of a year) recorded and located by OVSICORI-UNA.

We show that our pipeline detected a robust foreshock sequence, formed by ~23 earthquakes, that preceded the occurrence of the M6.5 mainshock, that was not observed and/or cataloged by traditional seismological routines at OVSICORI-UNA using the Antelope and Seiscomp3 software. These events present an average magnitude of 0.8 ± 0.3, based on relative amplitude measurements, and represent novel information about the nucleation phase of the mainshock.

Furthermore, we demonstrate that OKSP reproduced the manually located aftershock catalog from OVSICORI-UNA at the 90\% detection probability threshold and extend it with 1100 more earthquakes (at the 80\% detection probability threshold) that were not located by OVSICORI-UNA during the first week of the Puerto Armuelles earthquake sequence. 

The total number of new earthquakes detected by our pipeline fits well with previously detected earthquakes elsewhere using Artificial Intelligence (AI) algorithms, e.g. \cite{Baker, Tan2021} and represents the first stage for the automatic earthquake detection and location procedure in Costa Rica using AI systems.

\section{Computational Workflow}

\begin{figure*}[thbp]
    \centerline{\includegraphics[width=0.85\paperwidth]{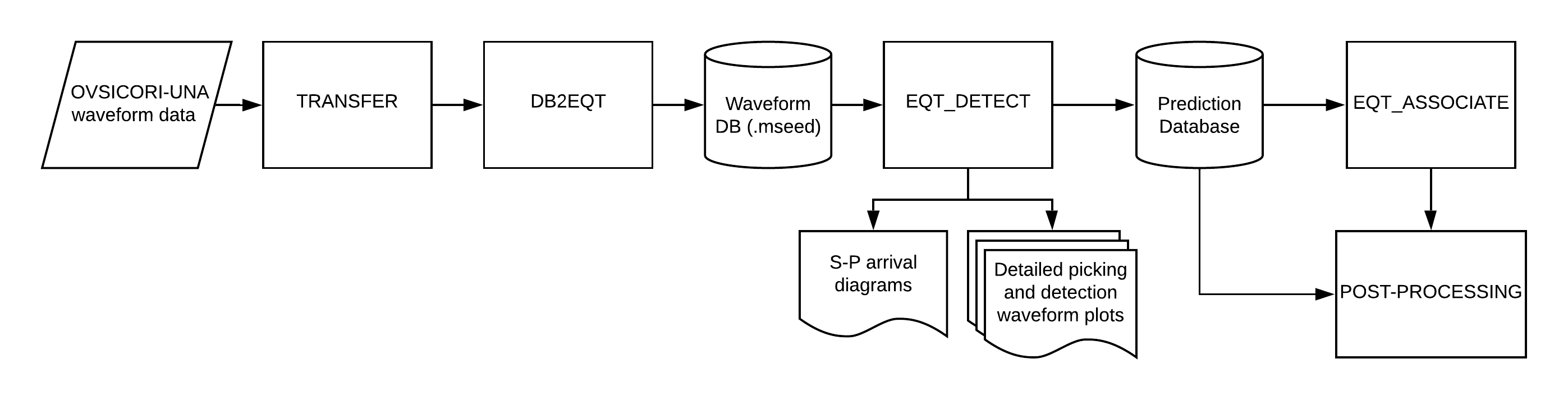}}
    \caption{\textbf{OKSP pipeline.} Flow diagram describing the earthquake detection and phase picking pipeline implemented at the Costa Rica High Technology Center (CeNAT) using three-component seismic data from OVSICORI-UNA
}
    \label{fig:fig3}
\end{figure*}

\subsection{EQTransformer}

For detecting new earthquakes, phase picking and comparing the earthquake detection capabilities of modern AI algorithms with those from traditional seismological methods, our pipeline incorporated the use of EQTransformer (hereafter EQT) a novel, open source, deep learning model designed for unsupervised recognition and phase picking of earthquakes \cite{Mousavi2020} from daily three-component seismic waveforms. This model was originally trained using the Stanford Earthquake Dataset (STEAD) \cite{Mousavi19}, a globally distributed training set consisting of 1.3 million local earthquake observations. The majority of these events are smaller than M2.5 and have been recorded within epicentral distances of 100 km. EQT was first tested with the aftershock sequence of 6 October 2000, M6.7 Tottori, Japan earthquake \cite{Mousavi2020}. We will briefly describe the network architecture of EQT next.

During the first stage, the multi-task structure of EQT consists of a very-deep encoder that converts the seismic signals from the time-domain into high-level representations of their temporal dependencies by using convolutional and LSTM (Long-Short Term Memories) blocks prior to the self-attention layers. LSTMs are a type of recurrent neural network (RNN) designed for handling large sequences. Its main element is a memory cell that, given an input, outputs a hidden state and updates the memory through a gate mechanism. Doing that avoids the problem of not learning long-term dependencies (that happens in practice in regular RNNs). The convolutional blocks exploit local structure and provide better temporal invariance, whereas the use of LSTM layers has proven to be important for incorporating positional information.

Next, an attention mechanism is implemented in a hierarchical structure, where a transformer performs a global attention along the full sequence, which is the direct input for the decoder of the detection predictor. Transformers use convolutional networks along with an internal self-attention mechanism for boosting the sequence transduction and learning long-term dependencies faster than regular recurrent or convolutional architectures \cite{vaswani}. The output of the global attention mechanism is also the input for the local attention blocks in the picking submodels. These local attention blocks examine small portions of the waveform and incorporate the information obtained by the global attention stage.

The resultant information is then used by three separate decoders to map three sequences of probabilities associated with 1) the existence of an earthquake signal (discriminating earthquakes from the background noise, anthropogenic or atmospheric sources), 2) the P-phase prediction probability (the probability that the picked phase corresponds to the arrival time of the P-wave)  and 3) the S-phase prediction probability (the probability that the picked phase corresponds to the arrival time of the S-wave). In order to assess the quality of the prediction probabilities, the decoders also provide output statistical variations based on Bayesian inference. 

\subsection{Automatic detection and phase picking}

The network structure of our pipeline is depicted in Fig. \ref{fig:fig3}. As input, we used 24-hr broadband, three-component waveforms from 10 seismic stations located in the study region and operated by OVSICORI-UNA (Fig. \ref{fig:fig2}). The \textit{TRANSFER} stage creates a waveform database for the selected group of stations and period of analysis. Then, in \textit{DB2EQT}, the seismic recordings in the database are converted to the file format accepted by EQT.

Once the database is in the proper format, \textit{EQT\_DETECT} performs the event detection and phase picking by doing a forward pass over an EQT trained model. For this purpose, EQT provides 2 models that were trained with different datasets. We selected the most up-to-date training model that incorporates the highest amount of earthquake observations worldwide with magnitudes below M2.5 \cite{Mousavi19} to minimize the false-positive rate of the predictions.

The main output from this stage consists of a prediction database for each station used in the analysis (Fig. \ref{fig:fig2}). The obtained database includes the origin times and event duration for all the earthquakes detected by the model, along with their corresponding detection probability and statistical variation. The P and S wave arrival times, and their associated prediction probabilities, is also included. 

At this stage we developed and applied quality-control (QC) routines that include the monitoring of time-domain waveforms for all the detected events at all the stations used, and Wadati differential travel time diagrams to assess P and S pick correctness and avoid false detections.

Finally, \textit{EQT\_ASSOCIATE} uses the prediction results and seismic network for performing the event association, removing outliers and building the final seismic catalog for the study period. The visualization and QC of the obtained seismic catalog are performed in the \textit{POST-PROCESSING} stage.

\subsection{Experimental setup}

\begin{table}[]
\caption{Specification of Computing Tools.}
\label{tab:tools}
\begin{tabular}{|l|l|l|}
\hline
\multicolumn{1}{|c|}{\textit{\textbf{Tool}}}                 & \multicolumn{1}{c|}{\textit{\textbf{Version}}} & \multicolumn{1}{c|}{\textit{\textbf{Description}}}                    \\ \hline
\begin{tabular}[c]{@{}l@{}}Linux\\ Distribution\end{tabular} & CentOS 7                                       & Operating system                                                      \\ \hline
Linux kernel                                                 & 3.10.0-1160.11.1.el7.x86\_64                   & Operating system core                                                 \\ \hline
CUDA                                                         & 11.0                                           & \begin{tabular}[c]{@{}l@{}}Framework for GPU\\ computing\end{tabular} \\ \hline
TensorFlow                                                   & 2.4.1                                          & Deep learning library                                                 \\ \hline
Python                                                       & Intel Python 3.5                               & Programming language                                                  \\ \hline
Matplotlib                                                   & 3.3.4                                          & Visualization library                                                 \\ \hline
EQTransformer                                                & 0.1.59                                         & \begin{tabular}[c]{@{}l@{}}Seismic signal\\ detection library\end{tabular} \\ \hline
\end{tabular}
\end{table}

The entire seismic detection pipeline was implemented on the Kabré Supercomputer at the Costa Rica National High Technology Center (CeNAT). Kabré is a heterogeneous computing cluster with different sections for each application area. The artificial intelligence section contains \textit{nukwa} nodes, each equipped with a graphical processing unit (GPU), with two different configurations. The first configuration features nodes each with an Intel Xeon E3-1225 v5 processor with 4 cores running at 3.30GHz, 16 GB of main memory, and an NVIDIA Tesla K40c GPU with 12 GB of memory. The second configuration presents nodes each with an Intel Xeon Silver 4214R processor with 24 cores running at 2.40GHz, 32 GB of main memory, and an NVIDIA Tesla V100 GPU with 32 GB of memory.

Several computational tools were used to build the seismic detection pipeline. Python scripts were written for data transformation during pre- or/and post-processing stages. Table \ref{tab:tools} summarizes the specification for all computational tools in our execution environment.

\section{Results and discussion}

\begin{figure*}[thbp]
    \centerline{\includegraphics[width=0.83\paperwidth, height=0.55\paperwidth]{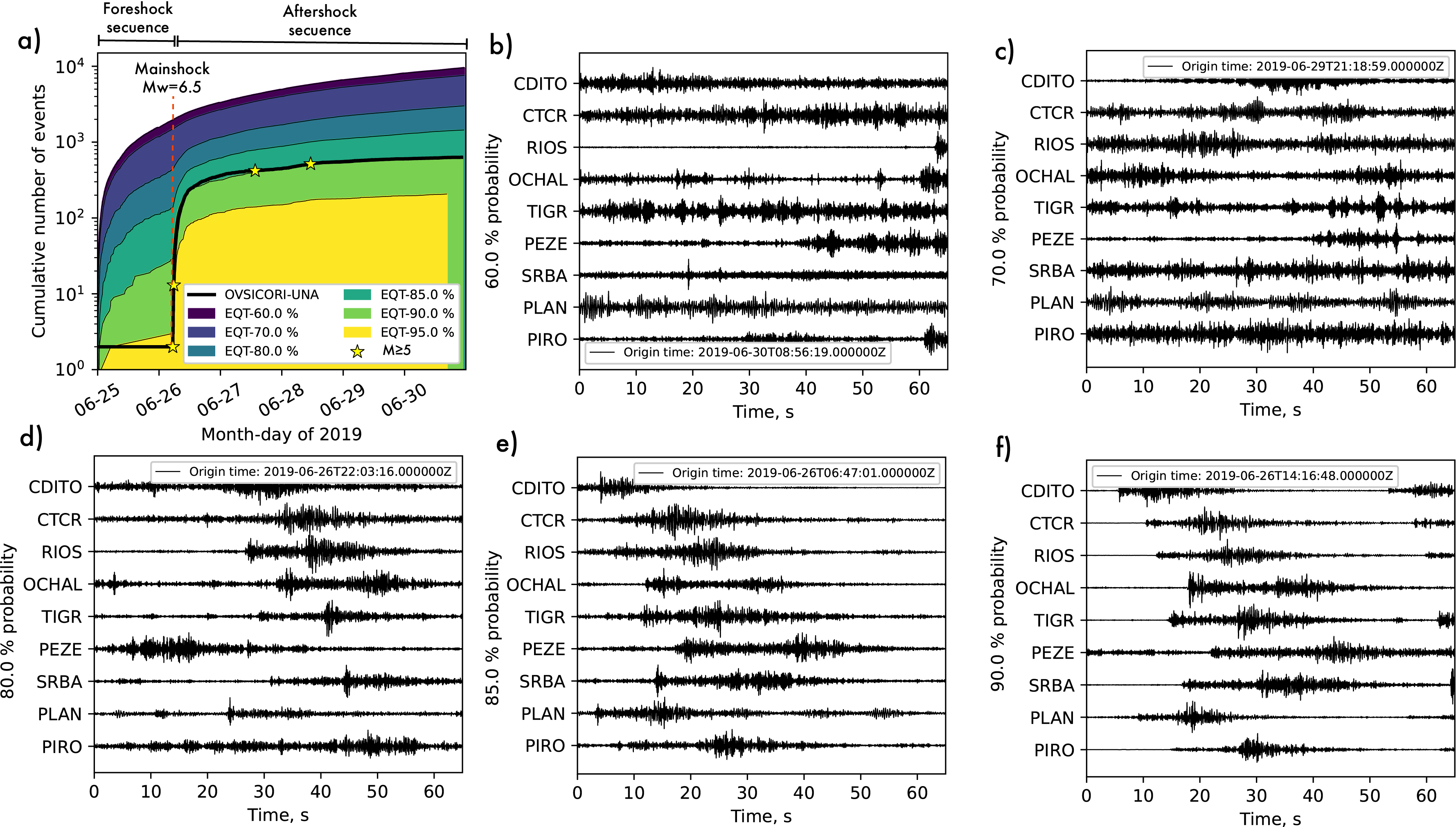}}
    \caption{\textbf{Figure showing the EQT seismic detection performance vs OVSICORI-UNA seismic catalog from June 25 to June 30, 2019.}  Panel a) shows the cumulative number (in logarithmic scale) of earthquakes detected by EQT as a function of time for the Puerto Armuelles earthquake sequence, color coded by the detection probability threshold. The black solid line represents the cumulative number of events located by OVSICORI-UNA for the same study period and region. The yellow stars on top of the black line highlight the occurrence of events with magnitudes  5.0. Panels b) though f) display recording examples of the vertical component for the stations shown in Fig. 2, extracted 20 s before and 70 s after the detected origin time and bandpass filtered between 1 and 15 Hz. The figure shows randomly selected events for each of the detection probability thresholds. 
}
    \label{fig:fig4}
\end{figure*}

We explored the detection and phase picking capabilities of EQT at different probability thresholds during our study period (Fig. \ref{fig:fig4}). Below the 80\% confidence interval, EQT is contaminated by an increasing number of “false positive” detections, due to the enhancement of ambient or/and anthropogenic noise (a reduction in the Signal/Noise ratio), as displayed in the vertical waveforms for all the stations used in this work (Fig. \ref{fig:fig4}b-\ref{fig:fig4}f). As a consequence, we are unable to separate pre-mainshock background seismicity rates from the observed aftershock sequence generated by the M6.5 earthquake.

Higher probability thresholds ($\geq$ 80\%) exhibited a significant improvement in the S/N ratio and highlighted an accelerated increase in the number of earthquakes immediately after the occurrence of the mainshock, as predicted by Gutenberg-Richter laws and found in previous studies in California and Italy (e.g. \cite{Ross, Tan2021}) using similar AI algorithms. As we are analyzing the first stage of the Puerto Armuelles earthquake sequence, our results do not show an Omori decay in the number of aftershocks \cite{Omori}. Based on the time-frequency behavior of the earthquake occurrence (Fig. \ref{fig:fig4}a) and S/N ratios exhibited in the waveforms at different detection probabilities, we selected 80\% as our lower detection bound for expanding the OVSICORI-UNA earthquake catalog for the Puerto Armuelles earthquake sequence.

\begin{figure}[bhtp]
    \centerline{\includegraphics[width=\columnwidth]{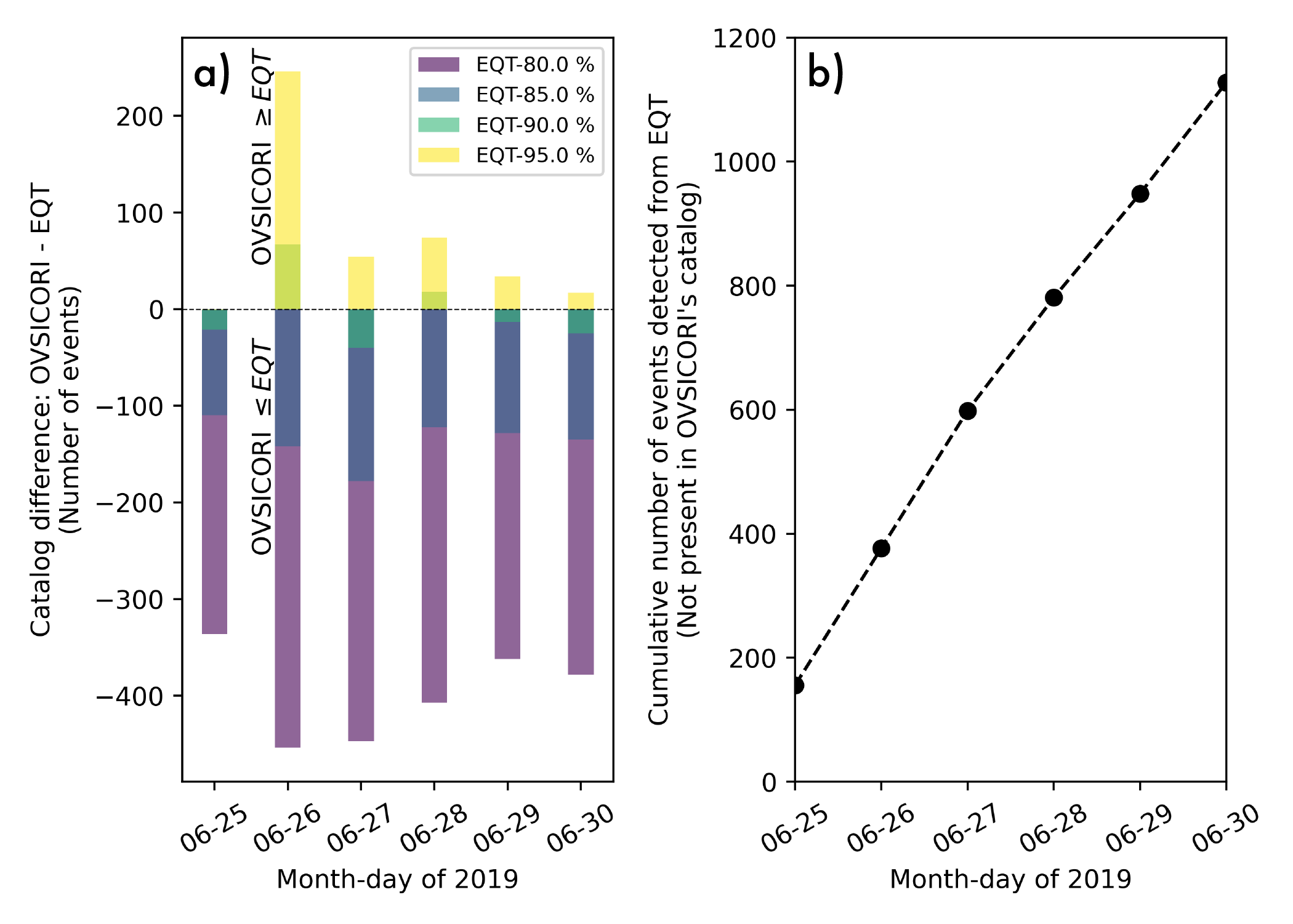}}
    \caption{\textbf{Comparison between the OVSICORI-UNA generated earthquake catalog and the AI-generated catalog through EQT and the OKSP pipeline.} Panel a) shows a histogram with the difference in the number of events detected by OVSICORI and EQT for the period of the analysis, color coded by detection probability. Negative values represent the number of earthquakes detected by EQT not present in OVSICORI’s catalog, whereas positive values correspond with events present in OVSICORI’s seismic catalog that were not detected by EQT at a particular probability threshold. The cumulative number of new events (nonexistent in the OVSICORI-UNA seismic catalog) as a function of time is shown in panel b).
}
    \label{fig:fig5}
\end{figure}

The difference per day in the number of earthquakes between EQT and OVSICORI-UNA, for different detection probabilities, is displayed in Fig. \ref{fig:fig5}a. In general, we found that for the 6 days of the analysis and detection probabilities between 80\% and 90\%, EQT outperformed the traditional detection capabilities developed by OVSICORI-UNA, and recovered a total of 1100 more earthquakes that were not detected and/or located previously by traditional approaches (Fig. \ref{fig:fig5}b).

\begin{table}[]
\centering
\caption{Classification metrics}
\label{tab:metrics}
\begin{tabularx}{0.8\columnwidth}{|X|X|}
\hline
\multicolumn{1}{|c|}{\textit{\textbf{Metric}}} & \multicolumn{1}{c|}{\textit{\textbf{Result}}} \\ \hline
Precision                                      & 0.8214                                        \\ \hline
Recall                                         & 1.0000                                        \\ \hline
F1 score                                       & 0.9020                                        \\ \hline
\end{tabularx}
\end{table}

Furthermore, at the 90\% threshold, EQT reproduced with incredible detail the mainshock-aftershock catalog generated by OVSICORI (shown as a black thick line in Fig. 4a) and expanded in 28 the number of new detections that anticipated the occurrence of the mainshock (potential foreshocks). We extracted and carefully reviewed the time-domain waveforms from all the pre-mainshock detections at the 90\% probability threshold and found that 23 out of the 28 events detected by EQT corresponded with real foreshocks that occurred from June 25 up to hours before the mainshock. These earthquakes were neither detected nor located by OVSICORI-UNA. For this period, missed detections of events already present in OVSICORI catalog (i.e false negatives) were not encountered, as it is shown in Fig. 5. Based on this information, the classification metrics for the model are computed and shown in Table \ref{tab:metrics}. We found that for this sample the model was 100\% exhaustive and 82\% precise, which gives an F1 score of ~0.90. These results are promising for a preliminary implementation and evaluation of our pipeline, as they show that it is possible to expand the OVSICORI-UNA seismic catalog in a robust manner once an adequate  probability threshold is established for the model. We computed the magnitude of the foreshock sequence through relative amplitude measurements and found an average magnitude of $0.8 \pm 0.3$.

Earthquakes with similar magnitudes and amplitudes are generally obscured in the time series by high frequency noise from natural sources (e.g. wind and ocean tides) and/or anthropogenic sources (e.g. diffuse, like highways, railways, etc., and harmonic from industrial machinery). Thus, recovering these small magnitude earthquakes is extremely important for improving our understanding of earthquake nucleation, rupture evolution and fault physics during the earthquake cycle.

\section{Concluding Remarks and Future Work}

This paper introduced OKSP, a novel pipeline based on deep learning algorithms for automatic earthquake detection in Costa Rica. Our results show that our framework is capable of efficiently processing waveform data from OVSICORI-UNA and accurately performing event detection and phase picking with an EQTransformer pretrained model. For our case study, the day before and the first 5 days following the Puerto Armuelles M6.5 earthquake that occurred on June 26,
2019, OKSP found 1100 more earthquakes previously unidentified by OVSICORI-UNA. The new data constitutes a robust sequence that allows to better understand the foreshock dynamic processes in the region.

Our experimental results demonstrate that AI algorithms, paired with high-end computer infrastructures and extensive databases, provide a state-of-the-art framework that may dramatically improve earthquake monitoring and understanding. OKSP is the first pipeline of this nature that has been proposed for earthquake detection in Costa Rica. OKSP expands current detection capabilities, reduces processing times and speeds up scientific quests in seismology. We see a promising future for OKSP in several dimensions. 

First, the proposed pipeline has the potential of speeding up the scientific response in seismic observation and monitoring, as we aim to make it capable of processing real-time waveform data from every available seismic stations in Costa Rica. Second, we also pursue the implementation of capabilities for massive data processing, which will help examine the extensive seismic information registered by OVSICORI-UNA over the past few decades. The insights that will be obtained from the expansion of the seismic catalog using novel methodologies, as the one we proposed, will contribute to the accurate communication and dissemination of information related to geodynamic processes of fault systems. Third, we acknowledge the boundless opportunity in exploring different deep learning models to improve detection and characterization of a wide variety of seismic phenomena. 

\section*{Acknowledgements}
This research was partially supported by a machine allocation on Kabré supercomputer at the Costa Rica National High Technology Center.

\bibliographystyle{IEEEtran}
\bibliography{references2}

\end{document}